\documentclass[aps,prl,amsmath,amssymb,amsfonts,showpacs,superscriptaddress,floatfix,twocolumn]{revtex4}

\usepackage{graphicx}
\usepackage{epsfig}
\usepackage[english]{babel}
\newcommand{\be}{\begin{equation}}
\newcommand{\bea}{\begin{eqnarray}}
\newcommand{\eea}{\end{eqnarray}}
\newcommand{\ee}{\end{equation}}

\newcommand{\calH}{\mathcal{H}}
\newcommand{\calJ}{\mathcal{J}}

\def\one{\ensuremath{\hbox{$\mathrm I$\kern-.6em$\mathrm 1$}}}

\begin{document}

\title{Lieb-Robinson bounds and the generation of correlations\\ and topological quantum order}
 \author{S. \surname{Bravyi}}
 \affiliation{IBM Watson Research Center, Yorktown Heights, NY 10598}

 \author{M. B. \surname{Hastings}}
 \affiliation{Center for Nonlinear Studies and Theoretical Division, Los Alamos National Laboratory, Los Alamos, NM, 87545}

 \author{F. \surname{Verstraete}}
 \affiliation{Institute for Quantum Information, Caltech, Pasadena, CA 91125}

\begin{abstract}
The  Lieb-Robinson bound states that local Hamiltonian evolution in nonrelativistic quantum mechanical theories
gives rise to the notion of an effective light-cone with exponentially decaying tails. We discuss several
consequences of this result in the context of quantum information theory. First, we show that the information
that leaks out to space-like separated regions is negligable, and that there is a finite speed at which
correlations and entanglement can be distributed. Second, we discuss how these ideas can be used to prove lower
bounds on the time it takes to convert states without topological quantum order to states with that property.
Finally, we show that the rate at which entropy can be created in a block of spins scales like the boundary of
that block.
\end{abstract}

\pacs{03.67.-a , 03.67.Mn, 03.65.Vf}

\maketitle

The principle of causality forms one of the pillars of modern physics. It dictates that there is a finite speed
at which information can propagate. Due to the existence of a light-cone, relativistic quantum field theories
automatically exhibit that property. The situation is however not so clear in non-relativistic quantum mechanics,
where a strict notion of a light cone does not exist. It has indeed been noticed that local operations can in
principle be used to send information over arbitrary distances in arbitrary small times \cite{lnc}. The seminal
work of Lieb and Robinson \cite{LR} and recent generalizations due to Hastings \cite{LRHas} and Nachtergaele and
Sims \cite{LRNach} however show that the situation in not so bad: if evolution is governed by local Hamiltonians,
then non-relativistic quantum mechanics gives rise to an effective light cone with exponentially decaying tails.
Due to this exponential attenuation, we will show how a quantitative version of causality emerges where the
amount of information that can be exchanged is exponentially small within space-time regions not connected by a
light-cone.

A related question is how fast correlations can be created between two widely separated regions in space. Note
that in this case, the principle of causality does not prohibit the build-up of correlations faster than the
speed of light, as correlations as such cannot be used to signal information; this is precisely the argument used
to show that the existence of entanglement does not violate causality. Again using the Lieb-Robinson bound, we
will show that there is a finite velocity at which correlations can be distributed. This automatically implies
that the time it takes to distribute entanglement between two nodes in a spin-network scales as the distance
between the nodes, solving an open question raised in \cite{Cubitt}. Note that we  assume that all classical
communication is also described by local Hamiltonian evolution, as otherwise it is possible to distribute
entanglement over arbitrary distances in a single unit of time by making use of the concept of quantum
teleportation \cite{repeater,localizable}.

Similar techniques can be used to prove lower bounds on the time it takes to create exotic quantum states
exhibiting topological quantum order \cite{Wen,Kitaev} by local Hamiltonian evolution: a time proportional to the
size of the system is needed. This is relevant in the light of topological quantum memories and computing, and
shows that the procedure described in \cite{topomem} to create toric code states is essentially optimal. Although
such a result is not too surprising on physical grounds, it is certainly nontrivial from the computational
complexity point of view as it is notoriously difficult to prove lower bounds on the depth of quantum circuits to
achieve specific tasks.

Finally, we will discuss bounds on the generation of entanglement by local Hamiltonian evolution: if a quantum
system is subject to a time-dependent Hamiltonian over a finite time, then an effective area law is obtained
which states that for large enough blocks, the increase of entropy of that block is at most proportional to its
surface. This is again relevant in the context of ground states of spin systems.

Let us start by define the kind of systems and evolution we will consider in this paper. We will consider a
spin-network endowed with a metric. For simplicity, let us assume that spins are located at vertices
of a graph $G=(V,E)$ and dynamics is generated by time-dependent
Hamiltonian  whose terms $h_{ij}(t)$ only couple nearest-neighbor spins:
\[
\mathcal{H}(t)=\sum_{(i,j)\in E }h_{ij}(t)
 \]
All results derived can however also be shown to hold in the case of fermions or local Hamiltonians with
exponentially decaying interactions, as the Lieb-Robinson bounds still apply.
Consider two  non-overlapping blocks of spins $A,B\subset V$.
We would like to know how operations in region $A$ affect observables in region $B$ at some later
moment of time. The Lieb-Robinson bound makes a statement about the operator norm
of the commutator of any operators $O_A$ and $O_B$
in regions $A,B$ taken at different
times; it states that
\[
\left\|\, \left[ O_A(t), O_B(0)\right]\, \right\|
\leq c N_{min}  \|O_A\|\, \|O_B\| \,
\exp\left(-\frac{L-v|t|}{\xi}\right),
\]
where $L$ is the distance between $A$ and $B$ (the number of edges in the shortest
path connecting $A$ and $B$), $N_{min}=\min\{ |A|, |B|\}$ is the number of vertices in
the smallest of $A$ and $B$,
 while $c,v,\xi>0$ are  constants~\cite{const}  depending only upon
$g=\max_{(i,j)\in E} \max_t \|h_{ij}(t)\|$ and maximum vertex degree of the graph (which we assume to be
constant)~\cite{tdep}.

Let us first check that this indeed bounds the amount of information that can be send
from $A$ to $B$ through the spin network. Let $C=V\setminus (A\cup B)$ be the part of the network
on which neither $A$ and $B$ have access to.  Without loss of generality, we can
assume that $A$ encodes her message by applying some unitary transformation $U^k_A$ on her subsystem where $k$ is
varied depending on the information she wants to send, i.e. a different unitary operation is applied if she wants
to send a different message (the most general operation she can implement is a completely positive map which can
indeed be implemented by unitary evolution with an extra ancilla which can be included in region $A$). Waiting for
time $t$, the whole system evolves according to the unitary operation
$U_{ABC}(t)$.
 If the global initial state of the system is given by $\rho_0$, then we can interpret this
procedure as a quantum channel where the input is
\[\rho^k_{ABC}=U^k_A\, \rho_0\, U^{k\dagger}_A\]
and the output \[\sigma^k_B(t)={\rm Tr}_{AC}\left(U_{ABC}(t)\,\rho^k_{ABC}\, U^\dagger_{ABC}(t)\right).\]
Let us show that $\sigma^k_B(t)$ has a very weak dependence on $k$. Indeed, denote
$\sigma^0_B(t)$
the state that $B$
would obtain if Alice would not have done anything (i.e. $U_A=\openone$).
Then for any observable $O_B$ acting on the subsystem $B$ and associated $O_B(t)=U^\dagger_{ABC}(t)O_BU_{ABC}(t)$, we have
\begin{eqnarray*}
{\rm Tr} \, \left( O_B\,(\sigma^0_B(t) - \sigma^k_B(t)) \right) &=& {\rm Tr} \,
\left( \rho_0\, U_A^{k\dagger}\,  [U_A^k,O_B(t)] \right) \\
 &\le &
 \|\, [U_A^k,O_B(t)]\, \| \le  \epsilon \, \|O_B\|,
\end{eqnarray*}
where $\epsilon$ is given by the
Lieb-Robinson bound:
\[\epsilon = c\, N_{min}\, \exp\left(-\frac{L-v|t|}{\xi}\right).\]
Therefore $\sigma^k_B(t)$ and $\sigma^0_B(t)$ are $\epsilon$-close in the trace norm:
\begin{equation}\forall k:
\|\sigma^k_B(t)-\sigma^0_B(t)\|_1 \leq \epsilon\label{ineq1}.
\end{equation}

If the probabilities to implement the unitaries $U_A^k$ are specified by $\{p_k\}$, then the amount of
information that is send through this quantum channel is given by the Holevo capacity:
\[
C_\chi(t)=S\left(\sum_k p_k\, \sigma^k_B(t)\right)-\sum_k p_k\, S\left(\sigma^k_B(t)\right)
\]
where $S(.)$ is the von-Neumann entropy.
Let $m$ be the Hilbert
space dimension of individual spins.
Combining equation (\ref{ineq1})
with the Fannes inequality
\[
|S(\eta_B)-S(\sigma_B)|\le \delta \, |B|\log m -\delta \log \delta, \quad \delta \equiv \| \eta_B-\sigma_B\|_1
\]
valid for any density operators $\eta_B$, $\sigma_B$ on the subsystem $B$,
we can bound the capacity as
$C_\chi(t)\leq 2\epsilon (|B|\log m - \log \epsilon )$.
Fix the time $t$ and increase the distance $L$, such that size of $B$
grows at most polynomially with $L$.
Clearly,  the capacity
$C_\chi(t)$ decreases exponentially fast with the distance
$L-v|t|$ and is hence negligible for distances $L\gg v|t|$. This indeed proves that the amount of
information that can be send outside the lightcone is exponentially small.

Let us next show that the amount of correlations that can be created by local Hamiltonian evolution vanishes also
exponentially outside an effective lightcone. Assume that we have a state $|\psi\rangle$ with finite correlation
length $\chi$, i.e. one in which all connected correlation functions,
$\langle O_A O_B \rangle_c\equiv \langle \psi|O_A(t)\, O_B(t)| \psi\rangle
- \langle \psi|O_A(t)|\psi\rangle \langle \psi|O_B(t)|\psi\rangle$,
 decay exponentially:
\[|\langle O_A\, O_B \rangle_c|
\leq
\tilde{c}\exp\left(-\frac{L}{\chi}\right),\]
for any regions $A$, $B$ with separation $L$, and any operators
 $O_A$, $O_B$ normalized such that $\|O_A\|,\|O_B\|\le 1$.
 The question we ask is the following: how long does it take to create
correlations between two regions separated by a distance $L$ when the evolution is generated by a local
Hamiltonian? For this purpose, we need the following ingredient. Consider an operator $O_A$ over region $A$
and the corresponding time-evolved operator $O_A(t)$.
We would like to prove that $O_A(t)$ can be well approximated by an operator acting on spins
{\it only} in the effective lightcone of $A$.
Choose an integer $l$ and let $S$ denote
the set of spins having distance at least $l$ from $A$. Denote
\[
O_A^l(t)=\frac{1}{{\rm Tr}_S (\openone_{S})}\,
{\rm Tr}_S (O_A(t))\otimes\openone_{S}
\]
Then the Lieb-Robinson bound allows us to prove that
\begin{equation}
\|O_A(t)-O_A^l(t)\|
\leq
c |A|  \exp{\left( -\frac{l-v|t|}{\xi}\right)},
\label{ineq2}
\end{equation}
Indeed,
let $U$ be a unitary operator acting on $S$ and $\mu(U)$ be the Haar measure
for $U$.  Then we have
\[
O_A^l(t)= \int d\mu(U) \, U\, O_A(t)\, U^\dag
\]
and therefore
\[
\begin{array}{r}\displaystyle
\|O_A(t)-O_A^l(t) \| \le \int d\mu(U)\,  \| \, [U,O_A(t)]\, \|
\end{array}
\]
Applying the Lieb-Robinson bound to the commutator
$[U, O_A(t)]$ with $N_{min}$ replaced by $|A|$
we arrive at Eq.~(\ref{ineq2}).

We now consider the connected correlation function
between $O_A$ and $O_B$ at time $t$.
Taking into account
$\|O_A^l(t)\|, \|O_B^l(t)\|\le 1$, we obtain
$|\langle O_A(t)\, O_B(t)\rangle_c|
\le
2c(|A|+|B|) \exp[-(l-vt)/\xi] +
|\langle O_A^l(t) O_B^l(t) \rangle_c|\le
2c(|A|+|B|) \exp[-(l-vt)/\xi] + \tilde{c} \exp[-(L-2l)/\chi]$.
Picking the optimal $l=(\chi vt + \xi
L)/(\chi+2\xi)$, we find that the connected correlation function at time $t$ is bounded by
$\bar{c} (|A|+|B|) \exp[-(L-2vt)/\chi'])$, where $\chi'=\chi+2\xi$,
hence proving that there is indeed a bounded velocity at which
correlations can be created.

As an application, let's consider the complexity of creating the GHZ-state \cite{GHZ}
$|\psi_{GHZ}\rangle=|00\cdots 0\rangle+|11\cdots 1\rangle$ out of the ferromagnetic state
$|\psi_{Fer}\rangle=|++\cdots +\rangle$ where all spins point in the $x$-direction; the spins are defined on an
arbitrary lattice (e.g. square lattice). Because the connected correlations of $|\psi_{GHZ}\rangle$ do not decay,
as opposed to the case of $|\psi_{Fer}\rangle$, the above results imply that the time it will take to transform
those states into each other by any time-dependent local Hamiltonian evolution scales linearly with the diameter
of the system. Typically such lower bounds on quantum circuits are difficult to prove, but in this case the
result follows directly from the Lieb-Robinson bounds.

Let's next apply the Lieb-Robinson techniques to a more exotic problem.  The concept of topological quantum order
(TQO) is exceptional in the sense that it is a property of a quantum state rather than of a Hamiltonian. Loosely
speaking, a quantum state $|\psi_1\rangle$ exhibits TQO if and only if there exists another
one orthogonal to it, $|\psi_2\rangle$, such that for all local observables $O_{loc}$ we have
$\langle\psi_1|O_{loc}|\psi_1\rangle=\langle\psi_2|O_{loc}|\psi_2\rangle$ and
$\langle\psi_1|O_{loc}|\psi_2\rangle=0$. TQO arises most frequently on systems with a
nontrivial topology, such as a torus. States with this property are natural candidates for protecting quantum
information from decoherence: decoherence can be thought of as a process in which an external quantum system
couples locally to the system of interest, and as such effectively acquires information about the state of the
system. However, if the quantum information is stored in a superposition of the two orthogonal states with
exactly the same local properties, then there is no way the environment can access that information, and if
$\langle\psi_1|O_{loc}|\psi_2\rangle=0$, there is no way the environment can correlate itself with it, and as
such cannot decohere it.

Formally, we define states $|\psi_1\rangle,|\psi_2\rangle$ to have TQO with error $(l,\epsilon)$ if
for any observable $O_{loc}$ with $\|O_{loc}\|=1$ supported on a set with diameter $l$ or less,
we have $|\langle \psi_1| O_{loc} |
\psi_1 \rangle- \langle \psi_2| O_{loc} | \psi_2 \rangle| \leq 2\epsilon$ and $|\langle \psi_1|
O_{loc}|\psi_2 \rangle|\leq \epsilon$. Colloquially, we say that a state is topologically ordered
if it has TQO to accuracy $(l,\epsilon)$ where $l$ is of order the linear size of the system, say
half the linear size, and $\epsilon$ is exponentially small in $l$, while a state is not toplogically ordered if
it has topological order only to accuracy $(l,\epsilon)$ with $l,\epsilon$ both of order unity.

We now ask the following question: starting from a state $|\psi_0\rangle$
which has no TQO,
is it possible to find lower bounds on the time it would take to create a state
$|\psi_1\rangle$ (or ``brother" state $|\psi_2\rangle$) with TQO if we allow for any local
time-dependent Hamiltonian evolution for a time $t$ with associated
unitary transformation $U$, so that $|\psi_1\rangle=U|\psi_0\rangle$?
We now prove by contradiction
that using the above colloquial definition of TQO,
it requires a time $t$ of order the linear size of the system to
achieve this.
Let the final state have TQO to accuracy $(l_f,\epsilon_f)$.
Consider now the state $|\tilde{\psi}_0\rangle=U^\dagger|\psi_2\rangle$.
Let $O_{loc}$ be a norm-$1$ operator with support on a set of diameter
$l_i$ (here and below the graph $G$ is a regular lattice in $\mathbb{R}^d$).
 We have, from Eq.~(\ref{ineq2}),
$|\langle\tilde{\psi}_0|O_{loc}|\tilde{\psi}_0\rangle-
\langle\psi_0|O_{loc}|\psi_0\rangle|=
|\langle\psi_2 | UO_{loc} U^{\dagger} | \psi_2\rangle-
\langle\psi_1 | UO_{loc} U^{\dagger} | \psi_1\rangle|
\leq {\cal}O(\epsilon_f+l_i^{d} \exp[-(l_f-l_i-vt)/\xi])$,
and similar bounds for matrix elements of $O_{loc}$ between $|\psi_0\rangle$
and $|\tilde{\psi}_0\rangle$.
Then, since $\langle\psi_0|\tilde{\psi}_0\rangle=0$, we can choose
$l_i=l_f/2$ in the above expression and it follows that
the initial state is topologically ordered to accuracy
$(\epsilon_i,l_i)$ with $\epsilon_i={\cal O}(\epsilon_f+l_f^{d-1}
\exp[-(l_f/2-vt)/\xi]$ and $l_i=l_f/2$.
Using the colloquial definition of TQO, then if
$vt<<l_f/2$, it follows that if the final state is topologically ordered,
so is the initial state, finishing the proof.
Again, the Lieb-Robinson bounds enable us to prove
lower bounds on the circuit complexity, in this case for creating states with TQO.
This proves, for example, that the strategy outlined in \cite{topomem} for creating toric code states is
essentially optimal.

Up till now, we have been  considering the evolution generated by local Hamiltonians, and made quantitative
statements about the speed at which correlations can be build up. In a similar vein, we can quantify how much
entanglement or entropy can be generated by such an evolution. In particular, we are interested in the amount of
entanglement that can be created  per unit of time between a block of spins $A$ and the rest of the system
$B=V\setminus A$.
We want to bound the rate at which entanglement is created by any Hamiltonian of the form
\[
\calH(t)=\calH_A(t)+\calH_B(t)+\sum_{k=1}^P r_k(t) \calJ_A^k\otimes \calJ_B^k\]
where operators $J_A^k$, $J_B^k$ act on their respective domains and we assume $\|J_A^k\|,
\|J_B^k\|\leq 1$.
Locality of $\calH(t)$ implies that the number of terms $P$ in the sum is proportional to the perimeter of $A$.
Any real-time evolution
can be approximated to arbitrary precision by a Trotter decomposition where at each time only one term
$\calJ_A^k\otimes \calJ_B^k$ couples $A,B$.
Now we  can use the results derived in \cite{entcap}
and later generalized in~\cite{CLV03}.
It was derived there
that the rate at which the entanglement, as measured by the entropy $S(\rho_A)$, can be created using any
product norm-$1$ Hamiltonian $\calJ_A\otimes \calJ_B$ is bounded above by a constant
\[c^*=2\max_{0\leq x\leq 1}\sqrt{x(1-x)}\log\frac{x}{1-x}\simeq 1.9.\]
This result is rather nontrivial because the maximum entangling rate does not depend upon dimensions of Hilbert
spaces describing $A$ and $B$ which may be arbitrarily large~\cite{imagtime}. Due to the Trotter approximation,
we have
\[\frac{dS(\rho_A)}{dt}\leq c^*\sum_{k=1}^P |r_k(t)|.\]
Hence the rate at which entanglement between $A$ and $B$ is created scales at most
like the perimeter of $A$ (and not as the volume).
If the interactions amplitudes are bounded by a constant,
$\max_k \max_t |r_k(t)|\le g$, we get
\[
S(\rho_A(t))-S(\rho_A(0))
\leq c^* \int_{\tau=0}^t d\tau \left(\sum_{k=1}^P |r_k(\tau)| \right)\leq c^* g P t.
\]
This is the result we were looking for: the amount of entanglement that can be created in a finite time in a
block of spins scales like the perimeter of the block and not as its volume. In particular, if one starts with a
state that obeys the area law, i.e. for which the entropy of large blocks scales like their perimeter, and
evolves it over some finite time, then it will still obey the area law.

This is relevant in the context of numerical renormalization group methods, as it is precisely the fact that
ground states obey an area law that leads to the remarkable precision of those methods \cite{DMRG}. Ground states
of spin systems belonging to the same phase can be converted into each other by local quasi-adiabatic evolution
\cite{quasi-ad} over a finite time, and it hence indicates that if there exists an efficient parametrization of
one state within a phase using matrix product states or generalizations, then all of them can be represented
efficiently.

In conclusion, we investigated apparent paradoxes arising in the context of causality and nonrelativistic
theories of quantum spin sytems. Lieb-Robinson bounds show that even in a nonrelativistic setting, a quantitative
notion of a lightcone arises where the light-cone has exponential tails which give rise to the apparent
paradoxes. The shape of the light cone is solely determined by the norm of the local terms in the Hamiltonian,
and is hence related to some speed of sound.  We have shown that the information that is leaking through the
exponential tails is useless, which shows that non-relativistic quantum mechanics is no-signalling in a
quantitative way. Similarly, we have shown that there is a finite speed at which correlations and entanglement
can be build up in a spin network. A nontrivial application of those results arises in the context of topological
quantum order: we could prove that the time it takes to create a state with topological quantum order out of one
that has not this property scales linearly in the size of the system. Finally, we have proven that the entropy of
a block of spins created by any local evolution coupling it to another domain scales at most like the surface of
the block.

{\it Acknowledgments---} S.B. was supported in part by National Science Foundation under Grant No. PHY99-07949,
MBH by US DOE W-7405-ENG-36, and FV by the Gordon and Betty Moore Foundation.


\begin{thebibliography}{99}
\bibitem{lnc} If the
Hamiltonian consists of a sum of local non-commuting terms, then $N$'th order perturbation theory will indeed
correlate observables at  sites separated by a distance $N$ even after an arbitrary small time.
\bibitem{const} More strictly, the Lieb-Robinson bound is given as $ \| [O_A(t),O_B(0)] \| \le 2 \|O_A\|  \|O_B\| F(L,t) $
where  $F(L,t)=\sum_{k=L}^\infty  1/k! (2g |t|)^k N(k) $ and $N(k)$ is the number of paths of length k which
connect $A$ and $B$. (in the case of a square lattice in the $D$-dimensions, a naive upper bound is given by
$N(k)\leq N_{min} (4D-1)^k$). The constants in the LR-bound are obtained by bounding $ F(L,t)\le N_{min} \exp{ ((
v |t| - L)/\xi ) } $. A choice of constants $\xi,v$ corresponds to a choice of a tangent surface bounding this
non-linear function from above.
\bibitem{tdep}
The standard proof of the Lieb-Robinson bound
applies to time-independent Hamiltonians. However the proof goes by setting up an upper
bound on the rate at which $\|\, [O_A(t),O_B(0)]\,\|$ changes with time.
Since we have a uniform upper bound on the interaction strength $\|h_{ij}(t)\|$,
the standard proof can be applied to the time-dependent Hamiltonians as well.
\bibitem{imagtime}
This result is not valid for imaginary time evolution, where the entanglement capacity of the Ising interaction
can be shown to be unbounded.
\end{thebibliography}
\end{document}